\documentclass[aip,jcp,reprint,superscriptaddress,longbibliography]{revtex4-2}%
\usepackage{graphicx}
\usepackage{color}
\usepackage{tabularx}
\usepackage{multirow}
\usepackage{dcolumn}
\usepackage{longtable}
\usepackage{tensor}
\usepackage[utf8]{inputenc}
\usepackage[english]{babel}
\usepackage[T1]{fontenc}
\usepackage{bm}
\usepackage{amsmath}
\usepackage{amsfonts}
\usepackage{amssymb}
\usepackage{adjustbox}
\usepackage{textcomp, gensymb}
\providecommand{\U}[1]{\protect\rule{.1in}{.1in}}

\newcommand{\blue}[1]{\textcolor{black}{#1}}

\def\be{\begin{equation}} %
\def\ee{\end{equation}} %
\newcommand{\bea}{\begin{eqnarray}}
\newcommand{\eea}{\end{eqnarray}}

\newcommand{\E}[2]{\hat{a}_{#1}^{\dagger} \hat{a}_{#2}}
\newcommand{\kappaQ}{\kappa_{\mathbf{Q}}}
\newcommand{\HF}{|\mathrm{HF}\rangle}
\begin{document}
\title{Probing Quantum Efficiency: Exploring System Hardness in Electronic Ground State Energy Estimation}
\author{Seonghoon Choi}
\affiliation{Department of Physical and Environmental Sciences, University of Toronto Scarborough, Toronto, Ontario M1C 1A4, Canada}
\affiliation{Chemical Physics Theory Group, Department of Chemistry, University of Toronto, Toronto, Ontario M5S 3H6, Canada}
\author{Ignacio Loaiza}
\affiliation{Department of Physical and Environmental Sciences, University of Toronto Scarborough, Toronto, Ontario M1C 1A4, Canada}
\affiliation{Chemical Physics Theory Group, Department of Chemistry, University of Toronto, Toronto, Ontario M5S 3H6, Canada}
\affiliation{Zapata Computing Canada Inc., Toronto, Ontario M5C 3A1, Canada}
\author{Robert A. Lang}
\affiliation{Department of Physical and Environmental Sciences, University of Toronto Scarborough, Toronto, Ontario M1C 1A4, Canada}
\affiliation{Chemical Physics Theory Group, Department of Chemistry, University of Toronto, Toronto, Ontario M5S 3H6, Canada}
\author{Luis A. Mart{\'i}nez-Mart{\'i}nez}
\affiliation{Department of Physical and Environmental Sciences, University of Toronto Scarborough, Toronto, Ontario M1C 1A4, Canada}
\affiliation{Chemical Physics Theory Group, Department of Chemistry, University of Toronto, Toronto, Ontario M5S 3H6, Canada}
\author{Artur F. Izmaylov}
\email{artur.izmaylov@utoronto.ca}
\affiliation{Department of Physical and Environmental Sciences, University of Toronto Scarborough, Toronto, Ontario M1C 1A4, Canada}
\affiliation{Chemical Physics Theory Group, Department of Chemistry, University of Toronto, Toronto, Ontario M5S 3H6, Canada}
\date{\today}
\begin{abstract}
We consider the question of how correlated the system hardness is between classical algorithms of electronic structure theory in ground state estimation and quantum algorithms. To define the system hardness for classical algorithms we employ empirical criterion based on the deviation of electronic energies produced by coupled cluster and configuration interaction methods from the exact ones along multiple bonds dissociation in a set of molecular systems. For quantum algorithms, we have selected the Variational Quantum Eigensolver (VQE) and Quantum Phase Estimation (QPE) methods. As characteristics of the system hardness for quantum methods, we analyzed circuit depths for the state preparation, the number of quantum measurements needed for the energy expectation value, and various cost characteristics for the Hamiltonian encodings via 
Trotter approximation and linear combination of unitaries (LCU).       
Our results show that the quantum resource requirements are mostly unaffected by classical hardness, with the only exception being the state preparation part, which contributes to both VQE and QPE algorithm costs. However, there are clear indications that constructing the initial state with a significant overlap with the true ground state (>10\%) is easier than obtaining the state with an energy expectation value within chemical precision. These results support optimism regarding the identification of a molecular system where a quantum algorithm excels over its classical counterpart, as quantum methods can maintain efficiency in classically challenging systems. 
\end{abstract}

\maketitle
\graphicspath{{./Figures/}}

\section{Introduction}
\label{sec:intro}

One of the main challenges faced by computational chemistry is solving the electronic structure problem, i.e., 
finding eigenstates of the molecular electronic Hamiltonian:
\begin{equation}
    \hat{H}_{e} = \sum_{pq=1}^{N} h_{pq} \E{p}{q} + \sum_{pqrs=1}^{N} g_{pqrs} \E{p}{q}\E{r}{s}, \label{eq:el_H}
\end{equation}
where $N$ is the number of spin-orbitals, and $h_{pq}$ and $g_{pqrs}$ are one- and two-body integrals.\cite{Helgaker:2013}
Even solving for the ground-state energy of $\hat{H}_{e}$ ($E_{0}$) exactly on a classical computer requires a computational cost that scales exponentially with the system size ($N$). Many classical computing algorithms\cite{DavidSherrill_Schaefer:1999, Crawford_Schaefer:2007, Chan_Sharma:2011, Baiardi_Reiher:2020} have been developed over the years. These algorithms can obtain $E_{0}$ with chemical precision, $\approx 1$ millihartree from the full configuration interaction (CI) answer within a fixed one-electron basis, with polynomial computational cost scaling with the system size for some systems. 
Yet, there are chemical systems that are still challenging for accurate description using classical algorithms, for example, many transition metal element complexes that are prospective homogeneous catalysts or organic light emitting diodes. The origin of difficulties in describing these compounds 
can be traced to presence of significant portions of dynamical and static electron correlation, this is also known as the strongly correlated regime. Without going to large compounds where the full CI answer is not available these complications can be modelled by considering small molecules with multiple chemical bonds partially broken. 
Partial bond breaking enhances the static electron correlation, and since the valence configurations are not decoupled from higher energy 
excited states, the dynamical part of electron correlation can be significant as well. 

Motivated by the recent developments in quantum computing hardware,\cite{IBM_roadmap:2020} quantum algorithms have surfaced as promising candidates for overcoming the exponential cost of solving the electronic structure problem.\cite{Abrams_Lloyd:1999, Kassal_Aspuru-Guzik:2008, Cao_Aspuru-Guzik:2019}  In the earlier phase of quantum algorithm development, the variational quantum eigensolver (VQE)\cite{Peruzzo_OBrien:2014,McClean_Aspuru-Guzik:2016,Ryabinkin_Izmaylov:2020,Cerezo_Coles:2021,Anand_Aspuru-Guzik:2022} was regarded as a promising hybrid quantum-classical algorithm for finding $E_{0}$ on noisy quantum computers of today. VQE invokes the variational principle to approximate $|\psi_{0}\rangle$ by finding a parameterized ansatz state with minimum energy. On a quantum computer, one prepares the parameterized ansatz state, $|\psi(\theta)\rangle$, and measures its energy, $E(\theta) = \langle \psi(\theta) | \hat{H}_{e} | \psi(\theta) \rangle$. Then, a classical optimizer is employed find optimal $\theta$ that minimizes $E(\theta)$. Because the efficiency of VQE depends strongly on the form of the ansatz, many quantum algorithms have been developed\cite{Ryabinkin_Izmaylov:2020, Tang_Economou:2021} to tackle this challenge. However, finding the appropriate $|\psi(\theta)\rangle$ state that can achieve chemical accuracy in large, strongly correlated systems still remains an open problem. Even estimating the scaling of quantum resources (e.g. circuit depth) with the system size poses a problem. 

While VQE algorithm in its current form may not be sufficient to solve the electronic structure problem completely, 
the solutions obtained by VQE (or classical algorithms inspired by VQE\cite{Ryabinkin_Genin:2021}) can be prepared on a quantum computer
as the initial guess ($|\phi\rangle$) for the subsequent quantum phase estimation (QPE) algorithm.\cite{Aspuru-Guzik_Head-Gordon:2005} 
Here, we assume availability of error corrected (fault-tolerant) hardware in the future. 
QPE can then obtain $E_{0}$ with chemical accuracy in polynomial time, provided that 
the VQE state has a non-negligible overlap with the exact ground state. QPE relies on the simple idea that the Fourier transformation of the autocorrelation function, $C(t) = \langle \phi | e^{-i \hat{H}_e t/\hbar} | \phi \rangle$, yields the spectrum, $S(E)$:\cite{Tannor:2007}
\begin{equation}
S(E) \equiv \frac{1}{2 \pi \hbar} \sum_{k} p_{k} \delta(E - E_{k}) =  \int_{-\infty}^{\infty} dt C(t) e^{i E t/\hbar},
\end{equation}
where $p_{k} \equiv |\langle \psi_{k} | \phi \rangle|^{2}$ is the overlap between the initial guess state and the $k$th eigenstate of $\hat{H}_e$ ($|\psi_{k}\rangle$). From $S(E)$, one can extract the $E_{0}$ energy. 

QPE, in its text-book form, has substantial circuit depth and ancilla qubit requirements. Consequently, to make QPE more suitable for early fault-tolerant quantum devices, several alternative versions still based on phase estimation have been developed.\cite{Griffiths_Niu:1996, Lin_Tong:2022, Wang_Johnson:2023, Ding_Lin:2023} These algorithms can find $E_{0}$ with $\epsilon$ accuracy with a single ancilla qubit, circuit depth of $\widetilde{\mathcal{O}}(\epsilon^{-1})$, and the total quantum run-time of $\widetilde{\mathcal{O}}(\epsilon^{-1}p_{0}^{-2})$. We use $\widetilde{\mathcal{O}}(\cdot)$ to denote $\mathcal{O}(\cdot)$ up to a poly-logarithmic factor. Yet, these favorable scalings are obtained assuming that the Hamiltonian evolution, $\hat{U}_H(t) \equiv e^{-i \hat{H}_e t/\hbar}$, can be implemented exactly on a quantum computer. In practice, one can only obtain $\hat{U}_H(t) |\phi \rangle$ approximately. Appendix~D of Ref.~\onlinecite{Lin_Tong:2022} shows that for a realistic estimate, both the circuit depth and the total run-time must be scaled by the cost of simulating $\hat{U}_H(\tau) |\phi \rangle$ on a quantum computer with $\mathcal{O}(\epsilon p_{0} \tau)$ accuracy. In this expression, to avoid aliasing of the spectrum, one has to choose $\tau = \mathcal{O}(\| \hat{H}_e \|_{\Delta}^{-1})$ with $\| \hat{H}_e \|_{\Delta}$ denoting the Hamiltonian spectral range, i.e., $\| \hat{H}_e \|_{\Delta} = E_{\mathrm{max}} - E_{0}$. 

The implementation of $\hat{U}_H(\tau)$ on a quantum computer can be achieved by employing either the linear combination of unitaries (LCU) method\cite{Childs_Wiebe:2012, Berry_Babbush:2019, Low_Wiebe:2019, Lee_Babbush:2021} or Trotterization.\cite{Suzuki:1990, Lloyd:1996, book_Hairer_Wanner:2006} The cost of QPE can thus be more accurately estimated by evaluating the cost of LCU and Trotterization. Evaluating the cost of these algorithms was extensively discussed in Refs.~\onlinecite{Martinez-Martinez_Izmaylov:2022, Loaiza_Izmaylov:2023, Loaiza_Izmaylov:2023a}. In these works, authors also suggest methods to reduce the cost of LCU and Trotterization and suggest the most efficient approach to implement $\hat{U}_H(\tau)$. \blue{Note that for LCU-based algorithms, performing phase estimation on the unitary $e^{i\arccos \hat H/\lambda}$ is more computationally efficient, where $\lambda$ represents the 1-norm of the LCU decomposition \cite{Babbush_Neven:2018}. This unitary is often referred to as the quantum walk unitary and can be implemented at a significantly lower cost than $\hat U_H(\tau)$ using the qubitization approach \cite{Low_2019}. The eigenvalues of $\hat H$ can then be easily recovered with straightforward classical post-processing. The computational complexity of LCU-based QPE primarily depends on $\lambda$, as well as the number and type of unitaries in the decomposition, regardless of which specific unitary is implemented. Therefore, for the sake of simplicity, we will focus our discussion on the time-evolution unitary $\hat U_H(\tau)$, keeping in mind that all discussed concepts can be applied directly to the walk-based QPE.}

The aim of the current work is to assess how the cost of the quantum algorithm for $E_{0}$ is affected by the classical hardness. Specifically, for several molecules in equilibrium, strongly correlated, and dissociated geometries, we compare the quantum cost of the two components of the quantum algorithm: VQE (for initial guess preparation) and the subsequent early fault-tolerant QPE.

We do not to present a detailed resource estimation of the quantum algorithm. Instead, the focus is on evaluating how the efficiencies of different components of the quantum algorithm are affected by the strong correlation, which poses a challenge for classical algorithms. 
We compare the cost of the following state-of-the-art quantum algorithms, needed for the ground-state energy estimation, at different molecular geometries: 
\begin{enumerate}
\item The accuracy of the initial guess obtained using the qubit coupled cluster (QCC) method.\cite{Ryabinkin_Izmaylov:2018, Ryabinkin_Izmaylov:2020, Ryabinkin_Genin:2021}
\item The number of quantum measurements and circuit depth required to obtain VQE energy: $E(\theta) = \langle \psi(\theta) | \hat{H}_{e} | \psi(\theta) \rangle$.
\item The cost of simulating $\hat{U}_{H}(\tau)|\phi\rangle$ using the LCU and Trotterization methods.
\end{enumerate}

In Sec.~\ref{sec:theory}, we present the metrics we use to evaluate points $1$--$3$ in more detail. Then, Sec.~\ref{sec:results} contains the numerical results for several molecular systems (H$_4$ linear chain, rectangular H$_4$, H$_2$O, and N$_2$) along with our discussion on the correlation between quantum and classical hardness. Finally, Sec.~\ref{sec:conclusion} concludes the paper.

\section{Theory}
\label{sec:theory}

\subsection{VQE ansatz built with QCC}
The total run-time of the early fault-tolerant QPE algorithms scale as $p_{0}^{-2}$, making it crucial to maximize $p_{0}$. As the system size increases, the overlap between the Hartree--Fock wavefunction ($|\mathrm{HF} \rangle$) and the exact ground state $|\psi_{0}\rangle$ is expected to diminish. Therefore, $\HF$ is typically no longer a valid initial guess state for the QPE algorithm.

To increase $p_{0}$ beyond that of $\HF$, one could employ one of the classical post-Hartree--Fock methods, such as CISD or coupled cluster singles and doubles (CCSD). However, the wavefunction resulting from these methods is difficult to prepare on a quantum computer. Instead, one can employ VQE to obtain a state with an appreciable overlap with $|\psi_{0}\rangle$ and, at the same time, easy to prepare on a quantum computer.

In VQE, the parameterized ansatz state has the following form:
\begin{equation}
|\psi(\theta)\rangle = \hat{U}(\theta)\HF.
\end{equation}
Therefore, on a quantum computer, one simply has to prepare the Hartree--Fock state,\cite{Google_HF:2020} then apply the parameterized unitary operation, $\hat{U}(\theta)$, to obtain $|\psi(\theta)\rangle$. Once the VQE is converged, we have the knowledge of the parameter that minimizes $E(\theta)$ (this optimal parameter is denoted as $\theta^{\ast}$). With the knowledge of $\theta^{\ast}$, it is simple to prepare $|\psi(\theta^{\ast})\rangle$ on a quantum computer for the subsequent QPE.

The performance of VQE strongly depends on the form of $\hat{U}(\theta)$. Among several VQE methods, each with a different $\hat{U}(\theta)$ form, we choose the QCC method.\cite{Ryabinkin_Izmaylov:2018} The QCC unitary can yield an accurate $|\psi(\theta^{\ast})\rangle$ with a moderate circuit requirement. The high performance of QCC can be attributed to the approach used in QCC to build $\hat{U}(\theta)$. In QCC, the unitary has the form
\begin{equation}
\hat{U}(\theta) = \prod_{k=1}^{N_{\mathrm{ent}}} e^{-i \theta_{k} \hat{P}_{k}},
\end{equation}
where $\hat{P}_{k}$ is a Pauli product: a tensor product of single-qubit Pauli and identity operators, i.e., $\hat{P}_{k} = \bigotimes_{n=1}^{N} \hat{\sigma}_{n}$ with $\hat{\sigma}_{n} \in \{\hat{x}_{n}, \hat{y}_{n}, \hat{z}_{n}, \hat{1}_{n}\}$. 

The basic idea of QCC is being frugal with $N_{\mathrm{ent}}$ by choosing $\hat{P}_{k}$ that has the largest potential to lower the energy. 
This minimization potential is approximated as
\begin{align}
\left.\frac{E(\theta_{k}; \hat{P}_{K})}{d \theta_{k}} \right\vert_{\theta_{k} = 0} &= \left.\frac{d \langle  \mathrm{HF} | e^{i \theta_{k} \hat{P}_{k}} \hat{H}_{q} e^{-i \theta_{k} \hat{P}_{k}} \HF}{d\theta_{k}} \right\vert_{\theta_{k} = 0} \nonumber \\
&= -i \langle \mathrm{HF} | [\hat{H}_q, \hat{P}_{k}] \HF,
\end{align}
where 
\begin{equation}
\hat{H}_{q} = \sum_{j} c_{j} \hat{P}_{j} \label{eq:qubit_H}
\end{equation}
is the qubit Hamiltonian obtained by applying one of fermion-qubit mappings (e.g., Jordan--Wigner or Bravyi--Kitaev transformation\cite{Bravyi_Kitaev:2002, Seeley_Love:2012}) to $\hat{H}_e$. We will use Jordan--Wigner transformation throughout this work. The Pauli products are ranked according to the size of the gradient at $\theta_{k} = 0$, and $N_{\mathrm{ent}}$ is kept small by only introducing $\hat{P}_{k}$'s with a large gradient. For further details on QCC, see Ref.~\onlinecite{Ryabinkin_Izmaylov:2018}, where the authors also discuss the secondary ranking based on the second derivative of $E(\theta_{k}; \hat{P}_{k})$ when the first derivative is zero.

In addition to a small $N_{\mathrm{ent}}$, each $e^{-i \theta_{k} \hat{P}_{k}}$ term in the QCC entanglers can be easily implemented on a quantum computer since it is an exponential of an $N$-qubit Pauli product. In Ref.~\onlinecite{Ryabinkin_Izmaylov:2018}, it was shown that the number of two-qubit unitaries grows as $N_{P_k}^{\log_{2} 3} \approx N_{P_k}^{1.585}$, where $N_{P_k}$ is the number of single-qubit Pauli operators in $\hat{P}_{k}$.

Yet another benefit of employing QCC is that there exists a version of QCC, called iterative QCC (iQCC), that can be efficiently run entirely on a classical computer.\cite{Ryabinkin_Genin:2021} Therefore, if one has only sufficient quantum resources for the main QPE, the guess state can be obtained using iQCC on a classical computer. In iQCC, instead of applying $\hat{U}(\theta)$ to $\HF$, it is applied to the Hamiltonian to obtain $\hat{H}_{\mathrm{eff}} = \hat{U}(\theta)^{\dagger} \hat{H}_{q} \hat{U}(\theta)$. Then, the minimum of $\langle \mathrm{HF} | \hat{H}_{\mathrm{eff}} \HF$ is found on a classical computer. In general, building $\hat{H}_{\mathrm{eff}}$ on a classical computer would be problematic due to the exponential growth of terms in the Baker--Campbell--Hausdorff (BCH) expansion of $\hat{U}(\theta)^{\dagger} \hat{H}_{q} \hat{U}(\theta)$. 
iQCC circumvents this issue by using the Pauli products in $\hat{U}(\theta)$, they satisfy $\hat{P}_{k}^2 = \hat{1}$, and thus $e^{i \theta_{k} \hat{P}_{k}} = \cos(\theta_{k}) + i \sin(\theta_{k}) \hat{P}_k$. This property makes construction of $\hat{H}_{\mathrm{eff}}$ efficient on a classical computer.
 Note that unlike other classical heuristics, the final VQE state obtained with iQCC on a classical computer can still be prepared easily on a quantum computer for the following QPE.

\subsection{VQE measurement cost}
\label{subsec:VQE_measurement}
In Ref.~\onlinecite{Gonthier_Romero:2022}, measuring $E(\theta) \equiv \langle \psi(\theta) | \hat{H}_{e} | \psi(\theta) \rangle$ on a quantum computer was identified as one of the bottlenecks of VQE. Measuring $E(\theta)$ is complicated because on a quantum computer, one can only measure the expectation value of polynomial functions of Pauli-$\hat{z}$ operators. 
However, the qubit Hamiltonian, $\hat{H}_{q}$, contains terms that are not all-$\hat{z}$. 

One of the most efficient methods for obtaining $E(\theta)$ starts by expressing the Hamiltonian as a sum of parts that can be easily rotated into a polynomial of Pauli-$\hat{z}$ operator:
\begin{equation}
\hat{H}_{q} = \sum_{\alpha} \hat{H}_{\alpha}. \label{eq:partitioned_H}
\end{equation}
To be more precise, $\hat{H}_{q}$ is partitioned such that for each $\hat{H}_{\alpha}$, one can easily find $\hat{U}_{\alpha}$ that rotates $\hat{H}_{\alpha}$ in to an all-$\hat{z}$ operator, i.e., 
\begin{equation}
\hat{U}_{\alpha} \hat{H}_{\alpha} \hat{U}_{\alpha}^{\dagger} = \sum_{p} c_{p}^{(\alpha)} \hat{z}_{p} + \sum_{pq} c_{pq}^{(\alpha)} \hat{z}_{p} \hat{z}_{q} + \dots \equiv p_{\alpha}(\hat{z}),
\end{equation}
where $\hat{z}_{p}$ is the Pauli-$\hat{z}$ operator acting on the $p$th qubit. Once the Hamiltonian is presented in the form in Eq.~(\ref{eq:partitioned_H}), one can then measure $E(\theta)$ by using
\begin{align}
\langle \psi(\theta) | \hat{H}_{q} | \psi(\theta) \rangle &= \sum_{\alpha} \langle \psi(\theta) | \hat{U}_{\alpha}^{\dagger} p_{\alpha}(\hat{z})\hat{U}_{\alpha} |  \psi(\theta) \rangle \nonumber \\
&= \sum_{\alpha} \langle \hat{U}_{\alpha} \psi(\theta) | p_{\alpha}(\hat{z}) | \hat{U}_{\alpha} \psi(\theta) \rangle. 
\end{align}

There exist numerous methods to partition the Hamiltonian into measurable $\hat{H}_{\alpha}$ fragments (see Ref.~\onlinecite{Choi_Izmaylov:2023a} for a summary). The quantum cost required to measure $E(\theta)$ depends significantly on the choice of the method for finding $\hat{H}_{\alpha}$. The partitioning method affects (1) the number of quantum measurements required to obtain $E(\theta)$ with $\epsilon$ accuracy,\cite{Yen_Izmaylov:2020, Crawford_Brierley:2021,Yen_Izmaylov:2022, Choi_Izmaylov:2022, Choi_Izmaylov:2023, Choi_Izmaylov:2023a} $M(\epsilon)$, and (2) the circuit requirements for implementing $\hat{U}_{\alpha}$'s on a quantum computer. To quantify (2), we use the average one- and two-qubit gate counts and the circuit depth required to implement $\hat{U}_{\alpha}$.

Out of many possible methods for Hamiltonian partitioning, we employ the fully commuting\cite{Yen_Izmaylov:2020} sorted insertion (FC-SI) method\cite{Crawford_Brierley:2021} and fluid fermionic fragment method applied to the low-rank decomposition (LR-F$^3$).\cite{Choi_Izmaylov:2023} FC-SI was chosen due to its practicality stemming from the balance between a low $M(\epsilon)$ and a low classical computational overhead. LR-F$^3$ was chosen since it has one of the lowest $M(\epsilon)$ among various partitioning techniques.\cite{Choi_Izmaylov:2023, Choi_Izmaylov:2023a} 

The FC-SI method partitions $\hat{P}_{j}$'s in $\hat{H}_q$ [Eq.~(\ref{eq:qubit_H})] into mutually commuting sets to find $\hat{H}_{\alpha}$. Each such set can be simultaneously rotated into all-$\hat{z}$ by an $N$-qubit Clifford rotation,\cite{Yen_Izmaylov:2020} which can be efficiently implemented on a quantum computer.\cite{Aaronson_Gottesman:2004} The greedy nature of the algorithm leads to a low $M(\epsilon)$. When the measurements are allocated optimally between the $\hat{H}_{\alpha}$'s, the total number of measurements is
\begin{equation}
M_{\mathrm{opt}}(\epsilon) = \frac{1}{\epsilon^2} \left[ \sum_{\alpha} \sqrt{\mathrm{Var}_{\psi}(\hat{H}_{\alpha})} \right]^2, \label{M_eps_opt}
\end{equation}
where $\mathrm{Var}_{\psi}(\hat{H}_{\alpha}) = \langle \psi(\theta) | \hat{H}_{\alpha}^2 | \psi(\theta) \rangle - \langle \psi(\theta) | \hat{H}_{\alpha} | \psi(\theta) \rangle^2$. The uneven distribution of variances arising naturally from the greedy nature of FC-SI leads to a smaller sum of square-roots arising in $M_{\mathrm{opt}}$ for a fixed $\sum_{\alpha} \mathrm{Var}_{\psi}(\hat{H}_{\alpha})$. However, the optimal measurement allocation requires the knowledge of $\mathrm{Var}_{\psi}(\hat{H}_{\alpha})$, which is unavailable. FC-SI circumvents this issue by approximating  $\mathrm{Var}_{\psi}(\hat{H}_{\alpha})$ with a classically efficient proxy for $|\psi(\theta)\rangle$, e.g., $\HF$ or $|\mathrm{CISD}\rangle$.

The LR-F$^3$ method first uses LR\cite{Huggins_Babbush:2021, Yen_Izmaylov:2021, Cohn_Parrish:2021} to decompose the molecular electronic Hamiltonian into 
\begin{align}
\hat{H}_e &= \hat{U}_{1}^{\dagger} \left( \sum_{p}^{N} \lambda_{p} \E{p}{p} \right) \hat{U}_{1} \nonumber \\ & \hspace{57pt} + \sum_{\alpha=2}^{N_{f}} \hat{U}_{\alpha}^{\dagger} \left( \sum_{pq}^{N} \lambda_{pq}^{(\alpha)} \E{p}{p} \E{q}{q} \right) \hat{U}_{\alpha}, \label{eq:frag_ham}
\end{align}
where $\hat{U}_{\alpha}$'s are orbital rotations, which diagonalize one-electron Hamiltonians,\cite{Yen_Izmaylov:2021} 
\begin{equation}
\hat{U}_{\alpha} = \exp\left[\sum_{pq} \gamma_{pq} (\E{p}{q} - \E{q}{p})\right].
\end{equation}
These unitaries can be efficiently implemented on a quantum computer by first decomposing them into a product of Givens rotations:\cite{Kivlichan_Babbush:2018}
\begin{equation}
\hat{U}_{\alpha} = \prod_{pq} \exp[\gamma_{pq}^{\prime} (\E{p}{q} - \E{q}{p})].
\end{equation}
Each term in Eq.~(\ref{eq:frag_ham}) is measurable because $\E{p}{p}$ maps to an all-$\hat{z}$ Pauli product under the usual fermion-qubit mapping, including Jordan--Wigner and Bravyi--Kitaev. The LR-F$^3$ method exploits the freedom in the LR fragments to optimize $M(\epsilon)$. Using the idempotency of $\E{p}{p}$, one can collect a fraction of the diagonal part of the two-electron $\hat{H}_{\alpha}$'s (with $\alpha > 1$) into a purely one-electron fragment:
\begin{align}
\hat{H}_{1}^{\prime} &= \hat{U}_{1}^{\dagger} \left( \sum_{p}^{N} \lambda_{p} \E{p}{p} \right) \hat{U}_{1} + \sum_{\alpha=2}^{N_{f}} \hat{U}_{\alpha}^{\dagger} \left( \sum_{p}^{N} c_{p}^{(\alpha)} \E{p}{p} \right) \hat{U}_{\alpha}, \nonumber \\
\hat{H}_{\alpha}^{\prime} &= \hat{U}_{\alpha}^{\dagger} \left[ \sum_{p}^{N} (\lambda_{pp}^{(\alpha)}-c_{p}^{(\alpha)}) \E{p}{p} + \sum_{p\neq q} \lambda_{pq}^{(\alpha)} \E{p}{p}\E{q}{q} \right] \hat{U}_{\alpha}, \label{eq:mod_te}
\end{align}
where $\alpha > 1$.
Because the collected one-electron $\hat{H}_{1}^{\prime}$ fragment is still a single one-electron term, it is easy to find a new unitary $\hat{U}_{1}^{\prime}$ that diagonalizes it. LR-F$^3$ minimizes $M(\epsilon)$ by optimizing $c_{p}^{(\alpha)}$'s. Like in FC-SI, because the exact variances are unavailable, one uses $M(\epsilon)$ approximated with a classically efficient proxy to optimize $c_{p}^{(\alpha)}$'s.

The difference in the circuit cost of FC-SI and LR-F$^3$ results from the difference in $\hat{U}_{\alpha}$ diagonalizing the $\hat{H}_{\alpha}$ fragments. Nevertheless, both FC-SI and LR-F$^3$ have favorable circuit costs required to implement corresponding $\hat{U}_{\alpha}'s$. The circuit depth requirements in both methods scale linearly with the system size ($N$). For FC-SI, the required two-qubit gate count to implement the $N$-qubit unitaries scales as $N^2/\log(N)$.\cite{Aaronson_Gottesman:2004} LR-F$^3$ is more demanding as it needs $N(N-1)/2$ two-qubit gates to implement $\hat{U}_{\alpha}$.\cite{Kivlichan_Babbush:2018} A low two-qubit gate count carries more significance than a low single-qubit gate count because the error rates (gate infidelities) of two-qubit gates are typically much higher.\cite{Arute_Martinis:2019, Bansingh_Izmaylov:2022}

\subsection{Cost of Hamiltonian simulation with Trotterization}
\label{subsec:Trotter}

Trotterization\cite{Suzuki:1990, Lloyd:1996, book_Hairer_Wanner:2006} implements $\hat{U}_{H}(t) \equiv e^{-i\hat{H}_{e}t/\hbar}$ as a product 
of exponential operators whose generators can be easily brought to the diagonal form. For example, in the first order Trotter formula,  
$\hat{U}_{H}(t)$ is approximated as $\prod_{\alpha} \hat{U}_{H_{\alpha}}(t)$, where
\begin{equation}
\hat{U}_{H_{\alpha}}(t) \equiv e^{-i\hat{H}_{\alpha}t/\hbar} = \hat{U}_{\alpha}^{\dagger} e^{-ip_{\alpha}(\hat{z}) t/\hbar} \hat{U}_{\alpha}, 
\end{equation}
and $\hat{H} = \sum_\alpha\hat{H}_{\alpha}$.
The exponentiation of an all-$\hat{z}$ operator ($p_{\alpha}(\hat{z})$) is trivial since an all-$\hat{z}$ operator is diagonal in
 the computational basis. Therefore,  implementing $\hat{U}_{H}(t)$ can be done using the same partitioning as in the 
measurement problem for VQE (see Sec.~\ref{subsec:VQE_measurement}). 

Unlike in the measurement problem where the sum of the fragment expectation values is exactly $E(\theta)$, the $\prod_{\alpha} \hat{U}_{H_{\alpha}}(t)$ product only approximately equals to $\hat{U}_{H}(t)$ because the $H_{\alpha}$ terms do not commute with each other. The error of this Trotterization is bounded from above by $\kappa t^{2} / 4$, where $\kappa$ is related to the spectral norms of the fragment commutators: $\kappa = \sum_{\alpha, \beta} \| [\hat{H}_{\alpha}, \hat{H}_{\beta} \|$.\cite{Martinez-Martinez_Izmaylov:2022} 

In practice, we are interested in the Trotter error of $\hat{U}_H(t)|\phi\rangle$, where $|\phi\rangle$ is the initial guess for QPE. In our scheme, $|\phi\rangle = |\psi(\theta^{\ast})\rangle$, i.e., the final converged VQE state. Exploiting the symmetries shared by every $\hat{H}_{\alpha}$, the Trotter error can be bound more tightly by replacing $\kappa$ with that evaluated in a symmetry subspace: $\kappaQ = \sum_{\alpha, \beta} \| [ \hat{H}_{\alpha}, \hat{H}_{\beta} ] \|_{\mathbf{Q}}$, where $\| \cdot \|_{\mathbf{Q}}$ denotes the spectral norm in the projected space pertaining to $\mathbf{Q}$. The subscript $\mathbf{Q}$ denotes the set of quantum numbers that define the manifold spanned by the states that lie in the same symmetry sector as $|\phi \rangle$. In chemistry, one typically knows \textit{a priori} the symmetry sector that $|\phi\rangle$ is in.\cite{Su_Campbell:2021} For example, we know the number of electrons, $\langle \phi | \hat{N}_e | \phi \rangle$, where $\hat{N}_e = \sum_{p} \E{p}{p}$.

Appendix~D of Ref.~\onlinecite{Lin_Tong:2022} shows that in the early fault-tolerant implementation of QPE, the Trotterization must approximate $\hat{U}_H(\tau)$ with $\mathcal{O}(\epsilon p_{0} \tau)$ accuracy, where $\tau = \pi \|\hat{H}\|_{\Delta}^{-1} / 3 \approx \| \hat{H} \|_{\Delta}^{-1}$. However, based on the typical $\kappaQ$ values from Ref.~\onlinecite{Martinez-Martinez_Izmaylov:2022}, the error of single-step Trotterization in approximating $\hat{U}_H(\tau)$ is much larger than $\epsilon p_{0} \tau$. To lower the Trotterization error to $\mathcal{O}(\epsilon p_{0} \tau)$, one can estimate $\hat{U}_H(\tau)$ using $N_{s}$ Trotter steps, each with a timestep of $\tau/N_{s}$. The error in this multi-step Trotterization can be bounded from above as:
\begin{equation}
\left\| U_{H}(\tau) - \left[\prod_{\alpha} \hat{U}_{H_{\alpha}}(\tau / N_{s})\right]^{N_{s}} \right\| \leq \kappaQ \tau / N_{s}.
\end{equation}

Choosing $N_{s} = \widetilde{\mathcal{O}}(\kappaQ \epsilon^{-1} p_{0}^{-1} \tau)$ ensures that the Trotterization error is sufficiently small for the QPE to yield $E_{0}$ with $\epsilon$ accuracy. With this consideration, the maximum circuit depth required in the early fault-tolerant QPE is $\widetilde{\mathcal{O}}(\kappaQ \epsilon^{-2} p_{0}^{-1})$ times the circuit depth for a single Trotter step of $\prod_{\alpha} \hat{U}_{H_{\alpha}}(\tau / N_{s})$. Similarly, the total run-time is $\widetilde{\mathcal{O}} (\kappaQ\epsilon^{-2} p_{0}^{-3})$ times the run-time of a single Trotter step. The cost of QPE due to Trotterization can thus be analyzed by computing $\kappaQ$.

Like for the measurement problem in VQE, there exist numerous partitioning methods to find the $\hat{H}_{\alpha}$ fragments. Among them, FC-SI\cite{Yen_Izmaylov:2020, Crawford_Brierley:2021} and LCU 1-norm optimized low-rank decomposition (LR-LCU)\cite{Martinez-Martinez_Izmaylov:2022} were chosen due to their small $\kappaQ$. 

While LR-LCU still starts from the LR fragments in Eq.~(\ref{eq:frag_ham}), these fragments are optimized differently than in LR-F$^3$. Due to the difficulties in the direct optimization of the Trotter error, authors in Ref.~\onlinecite{Martinez-Martinez_Izmaylov:2022} use that the LCU 1-norm [Eq.~(\ref{eq:ferm_LCU_norm}) in later Sec.~\ref{subsec:LCU}] is proportional to the upper-bound of Trotter error. Therefore, in LR-LCU, the LR fragments are modified such that the LCU 1-norm is reduced.

The symmetries that can be exploited to obtain $\kappaQ$ depends on the fragmentation technique. In LR-LCU, the number of electron symmetry ($\hat{N}_{e}$) is shared by every $\hat{H}_{\alpha}$ and the corresponding $\mathbf{Q}$ is the total number of electrons in the neutral molecule. In FC-SI, the Pauli products commuting with every $\hat{H}_{\alpha}$ ($\hat{P}_{i}$'s) can be found by using the qubit algebra (see Refs.~\onlinecite{Yen_Izmaylov:2020, Bravyi_Temme:2017, Setia_Whitfield:2020}). Then, the correct symmetry manifold can be determined by computing $\langle \mathrm{HF} | \hat{P}_{i} | \mathrm{HF} \rangle$, where  $| \mathrm{HF} \rangle$ is the Hartree--Fock approximation to $|\psi_{0}\rangle$.

\subsection{Cost of Hamiltonian simulation with LCU}
\label{subsec:LCU}

As an alternative to Trotterization one can use the LCU Hamiltonian encoding
\begin{equation}
\hat{H}_e = u_0 \hat{1} + \sum_{k=1}^{N_U} u_k \hat{U}_{k},
\end{equation}
where $u_k$'s are complex coefficients, and $\hat{U}_{k}$'s are unitary operators. 
Given the LCU decomposition of the Hamiltonian, on a quantum computer, 
one can construct a Hamiltonian oracle: a circuit that performs the action of $\hat{H}_e$ on $|\phi\rangle$.\cite{Childs_Wiebe:2012, Berry_Somma:2015, Low_Wiebe:2019, Loaiza_Izmaylov:2023, Loaiza_Izmaylov:2023a} 
This Hamiltonian oracle requires $\mathcal{O}[\log(N_U)]$ ancilla qubits to implement. 
The circuit requirements (gate count and depth) scale linearly with $N_U$; we refer the reader to Refs.~\onlinecite{Berry_Somma:2015,Babbush_Neven:2018} for further details on the Hamiltonian oracle circuit.

The number of calls to the Hamiltonian oracle required to obtain $\hat{U}_H(\tau)$ with desired accuracy in the LCU method scales as $\widetilde{\mathcal{O}}(\lambda \tau)$, where 
\begin{equation}
\lambda = \sum_{k=1}^{N_U} |u_{k}|
\end{equation}
is the LCU 1-norm.
Note that the desired accuracy only appears as a poly-logarithmic factor. Therefore, the early fault-tolerant QPE algorithm has a circuit depth that is $\widetilde{\mathcal{O}}(\lambda \epsilon^{-1})$ times that required for the Hamiltonian oracle. Similarly, the total run-time is $\widetilde{\mathcal{O}}(\lambda \epsilon^{-1} p_{0}^{-2})$ times the run-time of a single Hamiltonian oracle.\cite{Lin_Tong:2022}

References~\onlinecite{Loaiza_Izmaylov:2023, Loaiza_Izmaylov:2023a} discuss various methods for 
decomposing the Hamiltonian into an LCU. In this paper, we consider two of such methods: 
grouping anti-commuting Pauli products by employing sorted-insertion\cite{Crawford_Brierley:2021} (AC-SI) and the LR decomposition\cite{Huggins_Babbush:2021, Yen_Izmaylov:2021, Cohn_Parrish:2021} based method. 

Given a linear combination of mutually anti-commuting Pauli products:
\begin{equation}
\hat{A} = \sum_{a} c_{a} \hat{P}_{a},
\end{equation}
where $\{ \hat{P}_{a}, \hat{P}_{b} \} = 2 \delta_{ab} \hat{1}$, one finds that 
\begin{align}
\hat{A}^{\dagger} \hat{A} &= \sum_{ab} c_{a}^{\ast} c_{b} \hat{P}_{a} \hat{P}_{b} \nonumber \\
                          &= \sum_{a} |c_{a}|^2 + \sum_{a > b} c_{a}^{\ast} c_{b} \{\hat{P}_{a}, \hat{P}_{b} \}
                          &= \sum_{a} |c_{a}|^2,
\end{align}
making $\hat{A}/\sqrt{\sum_{a} |c_{a}|^2}$ a unitary operator. If the Hamiltonian can be partitioned into $N_U$ anti-commuting sets of Pauli products ($\hat{A}_{k} = \sum_{a} c_{a}^{(k)} \hat{P}_{a}$ for $k=1, \dots, N_U$), we have the following LCU decomposition
\begin{equation}
\hat{H}_q = \sum_{k=1}^{N_U} a_{k} (\hat{A}_{k}/a_{k}),
\end{equation}
where $a_{k} = \sqrt{\sum_{a} |c_{a}^{(k)}|^2}$, and the corresponding LCU 1-norm is $\lambda = \sum_{k} a_{k}$.

The LR-based decomposition first presents the Hamiltonian in the form of Eq.~(\ref{eq:frag_ham}) by using LR decomposition. 
One can turn this Hamiltonian into an LCU by replacing each $\E{p}{p}$ with $(\hat{r}_{p} + 1)/2$, where $\hat{r}_{p} \equiv 2 \E{p}{p} - 1$ is a unitary operator. Such replacement introduces additional terms in the one-electron part:\cite{Loaiza_Izmaylov:2023}
\begin{equation}
\hat{H}_{1}^{\prime} = \hat{U}_{1}^{\dagger} \left( \sum_{p}^{N} \lambda_{p} \E{p}{p} \right) \hat{U}_{1} + \sum_{pqr} g_{pqrr} \E{p}{q}.
\end{equation}
Like in LR-F$^3$, one can diagonalize this new one-electron Hamiltonian to express it as 
\begin{equation}
\hat{H}_{1}^{\prime} = \hat{U}_{1}^{\prime \dagger} \left( \sum_{p}^{N} \lambda_{p}^{\prime} \E{p}{p} \right) \hat{U}_{1}^{\prime},
\end{equation}
which can be turned into an LCU by the same $\E{p}{p} = (\hat{r}_{p} + 1)/2$ substitution. 
At the end of the entire procedure, we obtain:
\begin{align}
\hat{H}_e &= \frac{1}{2}\sum_{p} h_{pp} + \frac{3}{4}\sum_{pq} g_{ppqq} + \hat{U}_{1}^{\prime \dagger} \left( \sum_{p}^{N} \frac{\lambda_{p}^{\prime}}{2} \hat{r}_{p} \right) \hat{U}_{1}^{\prime} \nonumber \\ & \hspace{68pt} + \sum_{\alpha=2}^{N_{f}} \hat{U}_{\alpha}^{\dagger} \left( \sum_{pq}^{N} \frac{\lambda_{pq}^{(\alpha)}}{4} \hat{r}_{p} \hat{r}_{q} \right) \hat{U}_{\alpha}. \label{eq:frag_ham_2}
\end{align}
The constant factor can be neglected as it only introduces a constant phase in $\hat{U}_{H}(t)$.  
The resulting LCU 1-norm is 
\begin{equation}
\lambda = \sum_{p} \frac{|\lambda_{p}'|}{2} + \sum_{\alpha=2}^{N_{f}} \sum_{pq} \frac{|\lambda_{pq}^{(\alpha)}|}{4}. \label{eq:ferm_LCU_norm}
\end{equation}

Na{\"i}vely implementing each unitary term $\hat{r}_p$ and $\hat{r}_p \hat{r}_q$ in Eq.~(\ref{eq:frag_ham_2}) separately leads to an inefficient Hamiltonian oracle. Instead, for building the Hamiltonian oracle in LR, each fragment [labeled by $\alpha$'s in Eq.~(\ref{eq:frag_ham_2})] can be regarded as a single unitary because the \blue{complete square} structure of the fragment allows one to easily embed it into a unitary in a larger space.\cite{vonBurg_Troyer:2021, Lee_Babbush:2021} Therefore, the number of fragments ($N_f$) becomes a more important metric for estimating the circuit cost and ancilla requirements in the Hamiltonian oracle. However, it is important to note that embedding the fragment into a larger unitary will require additional ancillas. For a more detailed discussion, see Refs.~\onlinecite{vonBurg_Troyer:2021, Lee_Babbush:2021}.

As an extension of neglecting the constant term in Eq.~(\ref{eq:frag_ham_2}), one can use symmetry operators $\hat{S}$ satisfying $[\hat{H}_{e}, \hat{S}]$  to modify the Hamiltonian to decompose, thereby lowering the effective $\lambda$. One can exploit that the LCU 1-norm of the symmetry-shifted Hamiltonian $\hat{H}_e - \hat{S}$ is lower than that of the original $\hat{H}_{e}$. Due to the commutativity between $\hat{H}_e$ and $\hat{S}$, the following relationship is satisfied:
\begin{equation}
\hat{U}_{H}(t) = e^{-i (\hat{H}_e - \hat{S}) t/\hbar} e^{-i \hat{S} t/\hbar}.
\end{equation}
If $|\phi\rangle$ lies in the symmetry sector defined by an eigenvalue $s$, then 
\begin{equation}
\hat{U}_{H}(t) |\phi\rangle = e^{-i (\hat{H}_e - \hat{S}) t/\hbar} e^{-i s t/\hbar}|\phi\rangle.
\end{equation}
In Ref.~\onlinecite{Loaiza_Izmaylov:2023}, it was shown that the LCU 1-norms of symmetry-shifted Hamiltonians are 
much lower than those of the original Hamiltonians, achieving an almost two-fold reduction in $\lambda$. 

\section{Results and Discussion}
\label{sec:results}
\subsection{Choice of molecular systems}
Computational costs of different components of quantum algorithms discussed in Sec.~\ref{sec:theory} are each compared for molecules in equilibrium, strongly correlated, and dissociated geometries. The molecular systems we study are H$_4$ linear chain, H$_4$ in rectangular geometry (H$_4$ rect.), H$_2$O, and N$_2$ in the STO-3G basis. Owing to the small size of the molecules, the exact ground state can be readily calculated. The availability of this exact solution facilitates our analysis of classical and quantum hardness. 

When defining the equilibrium, strongly correlated, and dissociated geometries, we fix the bond angles and simultaneously stretch multiple bonds. As a result, the three different molecular geometries can be described by a single bond length. For H$_4$ linear chain, we extend all three H--H bonds simultaneously with the bond angle fixed at $\angle \mathrm{HHH} = 180\degree$. For H$_4$ rectangular, two of the parallel bonds were kept at $R(\mathrm{H} - \mathrm{H}) = 1$ \AA\ while the other two were stretched simultaneously. The H$_2$O bond angle was fixed at its experimental equilibrium value of $\angle \mathrm{HOH} = 104.5\degree$,\cite{Hoy_Bunker:1979} and both O--H bonds were simultaneously stretched. The single bond length parameter that describes the equilibrium, strongly correlated, and dissociated geometries for each molecule is summarized in Table~\ref{tab:geom}. Note that for H$_4$ rectangular, the molecule is the most stable when the two H$_2$ components are pulled far apart, i.e., the equilibrium coincides with the dissociated geometry.

\begin{table}[h!]
\caption{Bond lengths (in \AA ) defining equilibrium, strongly correlated, and dissociated geometries of the molecules used in our numerical study.}
\label{tab:geom}
\begin{ruledtabular}
\begin{tabular}{cccc}
Molecule  & Equilibrium & Correlated & Dissociated \\
\hline
H$_4$ chain &  0.9 &  2.0       &  3.0         \\
H$_4$ rect. & -     &  1.0         &  3.0         \\
H$_2$O &  1.0   &   2.1  &   3.0        \\
N$_2$ & 1.2  &   1.4      &   2.2        \\
\end{tabular}
\end{ruledtabular}
\end{table}

To determine the strongly correlated geometry, we use the CISD errors: $\epsilon_{\mathrm{CISD}} \equiv \langle \mathrm{CISD} | \hat{H}_e | \mathrm{CISD} \rangle - E_{0}$. The $\epsilon_{\mathrm{CISD}}$ values were evaluated at different geometries starting from $0.8$ \AA\ with $0.1$ \AA\ increment. The geometry with the maximum $\epsilon_{\mathrm{CISD}}$ was chosen as the strongly correlated geometry. Table~\ref{tab:CISDerror} shows that the $\epsilon_{\mathrm{CISD}}$ values are much larger at the strongly correlated geometry than those at the equilibrium and dissociated geometries. For H$_2$O and N$_2$, unrestricted Hartree--Fock reference was used for CISD while restricted Hartree--Fock was used for both types of H$_4$.

\begin{table}[h!]
\caption{CISD energy errors (in $10^{-3}$ a.u.) of different molecular systems used for our numerical analysis.}
\label{tab:CISDerror}
\begin{ruledtabular}
\begin{tabular}{cccc}
Molecule                    & Equilibrium           & Correlated      & Dissociated     \\  
\hline
H$_4$ chain                 &   0.766               & 43.1           & 8.40         \\
H$_4$ rect.                  & -                     & 4.64           & 2.04         \\
H$_2$O                      & 0.927                 & 78.9           & 0.245        \\
N$_2$                       & 28.1                  & 40.6           & 31.8         \\
\end{tabular}
\end{ruledtabular}
\end{table}

To further confirm that our definition of strongly correlated geometry is consistent with classical difficulty, we also report the errors in CCSD energies: $\epsilon_{\mathrm{CCSD}} \equiv \langle \mathrm{CCSD} | \hat{H}_e | \mathrm{CCSD} \rangle - E_{0}$ in Table~\ref{tab:CCSDerror}. The differences in $\epsilon_{\mathrm{CCSD}}$ values between the strongly correlated geometry and the other two geometries are even more profound than those for $\epsilon_{\mathrm{CISD}}$ values. The $\epsilon_{\mathrm{CCSD}}$ values thus validate that the strongly correlated geometry is indeed classically more challenging. Note that unrestricted Hartree--Fock reference was used for H$_2$O and N$_2$ like for CISD.

\begin{table}[h!]
\caption{CCSD energy errors (in $10^{-3}$ a.u.) of different molecules that were used for our numerical analysis.}
\label{tab:CCSDerror}
\begin{ruledtabular}
\begin{tabular}{cccc}
Molecule                       & Equilibrium     & Correlated      & Dissociated     \\  
\hline
H$_4$ chain                  & 0.007        & 3.72         & 0.114         \\
H$_4$ rect.                  & -               & 5.15         & 0.001         \\
H$_2$O                         & 0.140        & 25.4          & 0.282        \\
N$_2$                          & 4.19          & 9.68          & 1.98         \\
\end{tabular}
\end{ruledtabular}
\end{table}

\subsection{QCC energies and wavefunction overlaps}
\label{subsec:qcc_overlaps}

One strategy to solve the electronic structure problem on a quantum computer is to solely rely on VQE. In Table~\ref{tab:qcc_energy_overlaps}, we present the QCC energy errors for several molecules at classically easy (equilibrium and dissociated) and difficult (strongly correlated) geometries. Like in the classical methods (Tables~\ref{tab:CISDerror} and \ref{tab:CCSDerror}), unrestricted Hartree--Fock orbital basis was used for H$_2$O. However, we asses both the RHF and UHF orbital bases for the N$_2$ calculations. While the starting reference energies were expectedly lower in the UHF basis, the converged QCC energies and overlaps were improved when performed with RHF orbitals. In all QCC calculations, the reference state is kept fixed to the fermion-to-qubit mapped HF state, and the amplitudes were optimized using SciPy's \texttt{L-BFGS-B} implementation. To alleviate simulation cost, the H$_2$O calculations featured ffreezing of the oxygen 1$s$ orbital to be doubly occupied, resulting in a $12$ qubit Hamiltonian. Similarly, the N$_2$ calculations utilized a freezing of both nitrogen 1$s$ orbitals to be doubly occupied, resulting in a $16$ qubit Hamiltonian. The reported errors and overlaps for H$_2$O and N$_2$ are taken with respect the exact solutions in the frozen core approximation. The energy errors of the frozen core solutions compared to the solution in the unrestricted space were below $3 \times 10^{-4}$ a.u. for all instances. 

Table~\ref{tab:qcc_energy_overlaps} shows that quantum hardness faced by QCC to obtain accurate $E_{0}$'s is commensurate with classical hardness. In particular, a larger number of entanglers (i.e., greater quantum circuit resources) is required to obtain an accurate solution at strongly correlated geometries. In most molecules at strongly correlated geometry, chemical accuracy ($10^{-3}$ a.u. error) is not reached even with $N_{\mathrm{ent}} = 50$. Hence, the likelihood of QCC outperforming classical algorithms is low, given that the circuit cost of QCC becomes more demanding specifically for the types of systems where classical algorithms face challenges.

\begin{table}[h!]
\caption{Energy errors in $10^{-3}$ a.u. of converged QCC wavefunction with the exact ground state for varying number of generators $N_\mathrm{ent}$. The corresponding sum of overlaps of the QCC wavefunction with exact eigenstates within $1.5 \times 10^{-3}$ a.u. of $|\psi_{0}\rangle$ is in parentheses.}
\label{tab:qcc_energy_overlaps}
\begin{ruledtabular}
\begin{tabular}{c|c|ccc}
\multirow{2}{*}{Molecule}   & \multirow{2}{*}{Geometry}   & \multicolumn{3} {c}{$N_\mathrm{ent}$} \\ 
\cline{3-5}
 & & 10 & 20 & 50  \\  
\hline
\multirow{3}{*}{H$_4$ chain}   & Eq.  & 0.66 ($1.00$)      & 0.20 ($1.00$) & 0.14 ($1.00$)        \\
                               & Corr. & 6.0 ($0.975$)     & 3.2 ($0.984$) & 2.1 ($0.984$)  \\
                               & Diss.& 0.96 ($0.730$)     & 0.39 ($0.901$)  & 0.25 ($1.00$)  \\ \hline
\multirow{2}{*}{H$_4$ rect.}    & Corr. & 18 ($0.943$)     & 0.43 ($0.999$) & 4.4e-6 ($1.00$) \\
                               & Diss.& 0.34 ($1.00$)      & 2.4e-5 ($1.00$) & 4.8e-5 ($1.00$) \\\hline                         
\multirow{3}{*}{H$_2$O}  & Eq.  & 13 ($0.991$)            & 2.3 ($0.999$)            & 0.78 ($0.999$)     \\
                               & Corr. & 31 ($0.292$)            & 26 ($0.387$)           & 20 ($0.524$)   \\
                               & Diss.& 0.26 ($1.00$)           & 0.19 ($1.00$)           & 0.19 ($1.00$)    \\\hline
\multirow{3}{*}{{N$_2$ (RHF)}} & Eq.  & 47 ($0.970$)            & 25 ($0.981$)            & 6.6 ($0.997$)      \\
                               & Corr. & 106 ($0.875$)            & 76 ($0.919$)            & 34 ($0.970$)   \\
                               & Diss.& 275 ($0.000$)         & 271 ($0.00$)            & 127 ($0.548$)     \\\hline 
\multirow{3}{*}{{N$_2$ (UHF)}} & Eq.  & 93 ($0.781$)            & 79 ($0.787$)            & 64 ($0.791$)      \\
                               & Corr. & 114 ($0.405$)            & 92.3 ($0.497$)            & 80 ($0.517$)   \\
                               & Diss. & 7.8 ($0.266$)         & 6.9 ($0.271$)            & 6.7 ($0.281$)     \\
\end{tabular}
\end{ruledtabular}
\end{table}

Instead, one could use QCC as a method to obtain the initial guess state for QPE. Having a sizeable overlap between the guess state and $|\psi_{0}\rangle$ is important to reduce both the run-time and the circuit requirements of QPE (see Secs.~\ref{subsec:Trotter} and \ref{subsec:LCU}). If the target precision of the QPE ground state energy estimation is $\epsilon = 1.5 \times 10^{-3}$, the overlap with not only $|\psi_{0}\rangle$ is of relevance, but also the excited states possessing $E_k - E_0 \leq \epsilon$. Such states produce spectral peaks which merge into a single peak in QPE with $\epsilon$ resolution. Hence, in Table \ref{tab:qcc_energy_overlaps}, we include the sum of overlaps $\sum_{k} p_k$ of QCC with eigenstates $|\psi_{k} \rangle$ satisfying $E_k - E_0 \leq \epsilon$. 

With a relatively moderate number of entanglers ($N_{\mathrm{ent}}$ = 50), the QCC method can already obtain a quantum state having a sizeable overlap, $> 0.5$, with the exact ground state. This sizeable overlap allows one to perform a subsequent QPE (e.g., the method in Ref.~\onlinecite{Lin_Tong:2022}) to obtain $E_{0}$ with $\epsilon$ accuracy. The total run-time of this algorithm scales with $p_{0}^{-2}$, meaning that even the worst QCC states with $\sim 0.5$ overlap only slows down the QPE algorithm by a factor of $4$. 

Nevertheless, the smaller $p_{0}$ values for strongly correlated systems in QCC still suggest that the challenges faced by QCC are linked to classical hardness. This issue of increasing difficulty in finding a state with a sizeable $p_{0}$ as the system becomes larger and more strongly correlated was discussed extensively in Ref.~\onlinecite{Lee_Chan:2023}. The diminishing $p_{0}$ in QCC for classically difficult systems poses a genuine challenge for quantum algorithms aiming for practicality. Nonetheless, quantum heuristic algorithms and quantum-inspired algorithms, including QCC and iQCC, offer a distinct advantage over classical heuristics. Once we find a solution with a high $p_{0}$, this solution can easily be prepared on a quantum computer. 

\subsection{Quantum measurement cost in VQE}

In Tables~\ref{tab:si_meas} and \ref{tab:f3_meas}, we present the required number of preparations of $\hat{U}_{\alpha}|\psi(\theta)\rangle$ 
followed by its collapse onto the computational basis (i.e. measurements) for estimating $E(\theta) = \langle \psi(\theta)|\hat{H}_{e} |\psi(\theta)\rangle$ with $\epsilon$ accuracy in FC-SI and LR-F$^3$. To avoid estimating $M(\epsilon)$ at each VQE iteration, we use a configuration interaction singles and doubles wavefunction, $|\rm CISD\rangle$, to estimate $M(\epsilon)$ under the assumption that the converged VQE state will be similar to $|\rm CISD\rangle$. (In fact, to reduce the classical cost of computing $M(\epsilon)$, we truncate $|\rm CISD\rangle$ to only include as many Slater determinants as required to recover $99.99\%$ of 2-norm of $|\rm CISD\rangle$.) 
For quantum circuit estimates (1- and 2-qubit gate counts and circuit depth), $\hat{U}_{\alpha}$ were decomposed to 
exponentials of Pauli products\cite{Yen_Izmaylov:2020, Bansingh_Izmaylov:2022} 
(Givens rotations\cite{Kivlichan_Babbush:2018}) for FC-SI (LR-F$^3$) and then 
were converted to sequences of Hadamard, R$_{z}$, and CNOT gates using Qiskit.\cite{Qiskit}

\begin{table}[h!]
\caption{Quantum measurement cost in FC-SI to obtain Hamiltonian expectation value. We quantify the cost using the required number of measurements in millions to achieve $10^{-3}$ a.u. accuracy ($M$ for $\epsilon=10^{-3}$) and 1- and 2-qubit gate counts and circuit depth (averaged over different measurable fragments).}
\label{tab:si_meas}
\begin{ruledtabular}
\begin{tabular}{c|c|c|ccc}
Molecule                       & Geometry        & $M$    & 1-qubit   & 2-qubit  & depth \\  
\hline 
\multirow{3}{*}{H$_4$ chain} & Equilibrium     & 1.07            & 98              & 47             & 86\\
                               & Correlated      & 2.13            & 107             & 60             & 109 \\
                               & Dissociated     & 1.13            & 107             & 60             & 109 \\
\hline
\multirow{2}{*}{H$_4$ rect.} & Correlated      & 0.575           & 101             & 43             & 77 \\
                               & Dissociated     & 0.583           & 88              & 58             & 98 \\
\hline
\multirow{3}{*}{H$_2$O}        & Equilibrium     & 7.87            & 170             & 67             & 97\\
                               & Correlated      & 9.08            & 171             & 64             &96 \\
                               & Dissociated     & 0.657           & 179             & 66             & 101 \\
\hline
\multirow{3}{*}{N$_2$}         & Equilibrium     & 9.57            & 259             & 142            & 180\\
                               & Correlated      & 11.6            & 258             & 148            & 192\\
                               & Dissociated     & 5.21            & 276             & 152            & 199 \\
\end{tabular}
\end{ruledtabular}
\end{table}

\begin{table}[h!]
\caption{Quantum measurement cost in LR-F$^3$ to obtain Hamiltonian expectation value. We quantify the cost using the required number of measurements in millions to achieve $10^{-3}$ a.u. accuracy ($M$ for $\epsilon=10^{-3}$) and 1- and 2-qubit gate counts and circuit depth (averaged over different measurable fragments).}
\label{tab:f3_meas}
\begin{ruledtabular}
\begin{tabular}{c|c|c|ccc}
Molecule                       & Geometry        & $M$    & 1-qubit   & 2-qubit  & depth \\  
\hline  
\multirow{3}{*}{H$_4$ chain} & Equilibrium     & 0.595           & 97             & 48             & 62 \\
                               & Correlated      & 0.153           & 101             & 50            &  63 \\
                               & Dissociated     & 0.0147          & 106             & 52            & 67  \\
\hline
\multirow{2}{*}{H$_4$ rect.} & Correlated      & 0.538           & 101             & 50            & 65\\
                               & Dissociated     & 0.489           & 72             & 36             & 44 \\
\hline
\multirow{3}{*}{H$_2$O}        & Equilibrium     & 1.14            & 332             & 166           & 133  \\
                               & Correlated      & 0.206           & 328             & 164           &  134\\
                               & Dissociated     & 0.206           & 349             & 174           &  138 \\
\hline
\multirow{3}{*}{N$_2$}         & Equilibrium     & 2.90            & 673            & 336            & 196 \\
                               & Correlated      & 2.82            & 672            & 336            & 199 \\
                               & Dissociated     & 1.01            & 717            & 358            & 205  \\
\end{tabular}
\end{ruledtabular}
\end{table}

Both FC-SI and LR-F$^3$ exhibit little difference in the $M(\epsilon)$ values and circuit requirements between the three considered geometries. Moreover, for every molecule, the required number of measurements is lower for the correlated geometry than that for the equilibrium geometry in LR-F$^3$. The results suggest that, regarding the measurements in VQE, there is no correlation between quantum and classical hardness. Because quantum measurement is typically the bottleneck of VQE,\cite{Gonthier_Romero:2022} this similarity in measurement cost across different molecular geometries suggests that VQE is equally efficient regardless of classical hardness. Nevertheless, the quality of the solution obtained by VQE depends on $p_{0}$ reported in Sec.~\ref{subsec:qcc_overlaps}.

\subsection{Cost of Hamiltonian simulation in QPE}
The Hamiltonian simulation required for QPE can either be implemented using Trotterization or the LCU method. In Trotterization, the circuit depth and the total run-time of early fault-tolerant QPE scale linearly with $\kappaQ$. In the LCU method, they scale linearly with $\lambda$. 
\begin{table}[h!]
\caption{$\kappa_{\mathbf{Q}}$ in two Hamiltonian decomposition methods: FC-SI and LR-LCU. }
\label{tab:kappaq}
\begin{ruledtabular}
\begin{tabular}{c|c|cc}
Molecule & Geometry & FC-SI & LR-LCU\\
\hline 
\multirow{3}{*}{H$_{4}$ chain} & Equilibrium & 3.94 & 1.85 \\
 & Correlated & 1.39 & 0.34\\
 & Dissociated & 0.67 & 0.07\\
\hline 
\multirow{2}{*}{H$_{4}$ rect.} & Correlated & 1.20 & 1.59\\
 & Dissociated & 0.86 & 0.82\\
\hline 
 & Equilibrium & 118.10 & 49.34\\
H$_{2}$O & Correlated & 83.30 & 40.15\\
 & Dissociated & 43.28 & 38.85\\
\hline 
 & Equilibrium & 194.0 & 88.13\\
N$_{2}$ & Correlated & 176.37 & 83.4\\
 & Dissociated & 141.1 & 75.78\\
\end{tabular}
\end{ruledtabular}
\end{table}

\begin{table}
\caption{Spectral-range-distribution descriptors for the different systems
considered in this work. Here $C=\sum_{n}\Delta E_{n}$, where $\Delta E_{k}=E_{\max,k}-E_{\min,k}$
is the spectral range of the $k$th fragment $\hat{H}_{k}$ and $E_{\max,k}$
($E_{\min,k}$) is its maximum (minimum) eigenvalue. $S_{L}=1-\sum_{i}\omega_{i}^{2}$
, $\omega_{i}=\frac{\Delta E_{i}}{C}$ is the linearized entropy that
measures the degree of inhomogeneity in the distribution of spectral
ranges across a given decomposition scheme and $\beta=\sum_{i>j}\Delta E_{i}\Delta E_{j}=\frac{1}{2}C^{2}S_{L}\ge\sum_{i,j}||[\hat{H}_{i},\hat{H}_{j}]||_{\mathbf{Q}}$
is an upper bound to the (first-order) Trotter error. We consider
symmetry-projected Hamiltonian fragments akin to our $\kappa_{\mathbf{Q}}$
calculations in Table \ref{tab:kappaq}. }
\begin{centering}
\begin{tabular}{c|c|ccc|ccc}
\hline
\multirow{2}{*}{Molecule} &\multirow{2}{*}{Geometry}  & \multicolumn{3}{c}{FC SI} & \multicolumn{3}{c}{LR LCU}\\
 &  & $C$ & $S_{L}$ & $\beta$ & $C$ & $S_{L}$ & $\beta$\\
\hline
\multirow{3}{*}{H$_{4}$} & Eq. & 5.18 & 0.67 & 9.04 & 3.19 & 0.65 & 3.32\\
 & Corr. & 3.93 & 0.79 & 6.06 & 1.96 & 0.73 & 1.40\\
 & Diss. & 3.62 & 0.77 & 5.06 & 1.88 & 0.68 & 1.21\\
\hline
\multirow{2}{*}{H$_{4}$ rect} & Corr. & 3.93 & 0.55 & 4.21 & 2.57 & 0.71 & 2.35\\
 & Diss. & 3.65 & 0.63 & 4.19 & 2.81 & 0.71 & 2.81\\
\hline
\multirow{3}{*}{H$_{2}$O} & Eq. & 51.69 & 0.43 & 572.73 & 34.84 & 0.43 & 258.19\\
 & Corr. & 48.53 & 0.39 & 459.88 & 35.71 & 0.42 & 270.81\\
 & Diss. & 40.50 & 0.16 & 132.37 & 36.32 & 0.42 & 279.81\\
\hline
\multirow{3}{*}{N$_{2}$} & Eq. & 81.56 & 0.44 & 1469.59 & 50.74 & 0.50 & 648.62\\
 & Corr. & 79.92 & 0.45 & 1424.07 & 50.13 & 0.51 & 635.53\\
 & Diss. & 73.55 & 0.46 & 1246.80 & 49.25 & 0.51 & 618.52 \\
\hline
\end{tabular}
\par\end{centering}

\end{table}

\begin{table}[h!]
\caption{LCU 1-norm ($\lambda$) and the number of LCU unitaries ($N_U$) in AC-SI and number of fermionic fragments ($N_{f}$) in LR. \blue{The spectral range $\Delta E/2 \equiv (E_{\max} - E_{\min})/2$ corresponds to a lower bound for $\lambda$, for $E_{\max(\min)}$ the maximum(minimum) eigenvalue of the Hamiltonian.}}
\label{tab:lcu_unshifted_cost}
\begin{ruledtabular}
\begin{tabular}{c|c|c|cc|cc}
\multirow{2}{*}{Molecule} & \multirow{2}{*}{Geometry} & \multirow{2}{*}{$\Delta E/2$} & \multicolumn{2}{c}{AC-SI}  & \multicolumn{2}{c}{LR}  \\ 
 &  &  &  $\lambda$ & $N_U$ & $\lambda$  & $N_{f}$  \\ \hline
\multirow{3}{*}{H$_4$ chain} & Equilibrium & 2.85 & 5.05 & 62 & 5.14 & 18 \\ 
  & Correlated & 1.52 & 2.74 & 72 & 3.37 & 17 \\ 
  & Dissociated & 1.32 & 2.36 & 64 & 3.28 & 15  \\ \hline
  \multirow{2}{*}{H$_4$ rect.} & Correlated & 2.58 & 4.58 & 64 & 4.66 & 18 \\ 
  & Dissociated & 1.97 & 3.77 & 52 & 4.42 & 18 \\ \hline
\multirow{3}{*}{H$_2$O} & Equilibrium & 41.9 & 57.3 & 236 & 53.7 & 42  \\ 
  & Correlated & 39.5 & 53.6 & 238 & 50.5 & 41 \\ 
  & Dissociated & 38.8 & 50.7 & 320 & 49.3 & 39 \\ \hline
\multirow{3}{*}{N$_2$} & Equilibrium & 64.5 & 90.9 & 550 & 91.1 & 73 \\ 
  & Correlated & 63.0 & 89.2 & 552 & 89.8 & 71  \\ 
  & Dissociated & 59.6 & 84.1 & 948 & 87.0 & 67  \\ 
\end{tabular}
\end{ruledtabular}
\end{table}

\begin{table}[h!]
\caption{\blue{Same as Table \ref{tab:lcu_unshifted_cost} but for symmetry-shifted Hamiltonians.}}
\label{tab:lcu_cost}
\begin{ruledtabular}
\begin{tabular}{c|c|c|cc|cc}
\multirow{2}{*}{Molecule} & \multirow{2}{*}{Geometry} & \multirow{2}{*}{$\Delta E/2$} & \multicolumn{2}{c}{AC-SI}  & \multicolumn{2}{c}{LR}  \\ 
 &  &  &  $\lambda$ & $N_U$ & $\lambda$  & $N_{f}$  \\ \hline
\multirow{3}{*}{H$_4$ chain} & Equilibrium & 1.73 & 2.96 & 56 & 3.25 & 18 \\ 
  & Correlated & 0.78 & 1.69 & 58 & 2.02 & 17 \\ 
  & Dissociated & 0.75 & 1.63 & 58 & 1.90 & 16  \\ \hline
  \multirow{2}{*}{H$_4$ rect.} & Correlated & 1.26 & 2.21 & 62 & 2.33 & 14 \\ 
  & Dissociated & 1.42 & 2.58 & 52 & 2.81 & 16 \\ \hline
\multirow{3}{*}{H$_2$O} & Equilibrium & 23.8 & 27.9 & 228 & 27.6 & 40  \\ 
  & Correlated & 23.5 & 27.0 & 236 & 26.6 & 40 \\ 
  & Dissociated & 23.6 & 26.1 & 312 & 26.5 & 36 \\ \hline
\multirow{3}{*}{N$_2$} & Equilibrium & 35.0 & 45.1 & 558 & 47.6 & 69 \\ 
  & Correlated & 34.9 & 45.6 & 540 & 47.7 & 66  \\ 
  & Dissociated & 35.9 & 47.7 & 934 & 48.6 & 67  \\ 
\end{tabular}
\end{ruledtabular}
\end{table}

Tables~\ref{tab:kappaq}, \ref{tab:lcu_unshifted_cost}, and \ref{tab:lcu_cost} show that the $\kappaQ$ and $\lambda$ values are not correlated with classical hardness. Therefore, like the quantum measurement cost in VQE, the resource requirements for Hamiltonian simulation in QPE do not increase due to classical hardness. In fact, for most molecules, the $\kappaQ$ and $\lambda$ values are lower at strongly correlated geometry than those at the equilibrium geometry, making the quantum cost lower for the classically harder systems. 

The circuit cost and ancillas required for the Hamiltonian oracle in the LCU method are dictated by the number of unitaries ($N_U$) in AC-SI and the number of fragments ($N_{f}$) in LR. Both $N_U$ (in AC-SI) and $N_{f}$ (in LR) show a similar trend as $\lambda$ across different molecular geometries \blue{for both the full and symmetry-shifted Hamiltonians}, further supporting that the quantum hardness is not correlated with the classical hardness. 

\section{Conclusion}
\label{sec:conclusion}
This work studies the relationship between the hardness in classical algorithms and that in the VQE and QPE quantum algorithms for finding the ground-state energy of molecular systems. To survey systems that pose varying degrees of difficulties to classical algorithms, we consider several molecules at equilibrium, partially bond-broken (strongly correlated), and dissociated geometries. The strongly correlated geometry is challenging for classical systems to obtain the ground-state energy with high accuracy. 

We assessed the efficiency of two components of the quantum algorithms: initial guess state preparation and the subsequent algorithm for finding the ground-state energy. We found that the efficiency of components of the quantum algorithms is unaffected by the classical hardness except for the quality of the initial guess state. The initial guess state, which we obtained using the variational quantum eigensolver, had a lower overlap with the exact eigenstate at the strongly correlated geometry. But, even this low overlap was sufficiently sizeable to allow for the main quantum algorithm to find the ground-state energy (with $\approx 4$ times longer total run-time). We recognize that with growing system size, finding an initial guess state with sufficient overlap can become more difficult. 
Yet, based on our examples, it is clear that obtaining an initial state with a sizeable overlap requires lower ansatz complexity (i.e. the number of iQCC generators) than that to achieve chemical precision in energy.

While this study showed that classical hardness is mostly unrelated to quantum hardness, studying the feasibility through a detailed resource estimation of the quantum algorithm was outside the scope of this work. Nevertheless, this work shows that there remains some potential to find a molecular system where a quantum algorithm is more appropriate than classical counterparts, as the efficiency of these two types of methods is mostly disjointed.

\section*{Acknowledgments}
A.F.I. acknowledges financial support from the Natural Sciences and Engineering Council of Canada (NSERC) and the Defense Advanced 
Research Projects Agency (DARPA) within the Quantum Benchmarking Program. A.F.I. and I.L. gratefully appreciate financial support from Mitacs Elevate Postdoctoral Fellowship and Zapata Computing Inc. S.C. acknowledges financial support from the Swiss National Science Foundation through the Postdoc Mobility Fellowship (Grant No. P500PN-206649). This research was enabled in part by support provided by Compute Ontario and Compute Canada.

%

\end{document}